\documentclass[aps,prd,groupedaddress,showpacs,preprint]{revtex4} 
\usepackage{bm}
\usepackage[cp866]{inputenc}
\usepackage[english]{babel}

\usepackage{graphicx}

\input epsf

\begin{document}

\title{Study of Light Scalars,  the learned  Lessons}
\author{N.~N.~Achasov}
\email[]{achasov@math.nsc.ru}\affiliation{Laboratory of
Theoretical Physics, S.L. Sobolev Institute for Mathematics  SB
RAS, 630090, Novosibirsk, Russia}
\begin{abstract}
   Attention
is paid to the production mechanisms of the light scalars that
reveal their nature.\\[6pt]
 In   the linear sigma model it
is revealed the chiral shielding of the $\sigma(600)$ meson and
shown that the $\sigma$ field is described by its four-quark
component.\\[6pt]
   The  $\pi\pi$  scattering amplitude is constructed taking into account the $\sigma(600)$  and
 $f_0(980)$  mesons,  the chiral
shielding of  $\sigma (600)$, the $\sigma(600)$-$f_0(980)$ mixing,
 and results, obtained on the base of the chiral expansion and
the Roy equations. The data agree with the four-quark nature of
$\sigma (600)$ and $f_0(980)$.\\[6pt]
 It is shown that the kaon loop mechanism of the $\phi$ radiative decays into the light scalar mesons, which is ratified
by experiment, is the four-quark transition and points to the
four-quark nature of the light scalars.\\[6pt]
 It is shown also, that  the  light scalars
are produced in the two photon collisions via  four-quark
transitions in contrast to the classic $P$ wave tensor $q\bar q$
mesons, which are produced via two-quark transitions
 $\gamma\gamma\to q\bar q$, that points to the four-quark nature of the light scalar mesons,
 too.\\[6pt]
 A programme of
further investigations is laid down.
\end{abstract}
\maketitle

\section{Introduction \cite{GML}}

 Arising  50 years ago from the linear sigma
model ( LSM) \cite{GML},
 the light scalar meson problem  became central in
 the nonperturbative   QCD for  LSM could be its low energy
realization. The scalar channels in the region up to 1 GeV is  a
stumbling block of  QCD.  The point is that not only perturbation
theory fails here, but sum rules as well in view of the fact that
isolated resonances are absent in this region.

\section{QCD, chiral limit, confinement, \boldmath{$\sigma$} models }

  $L=-(1/2)Tr\left (G_{\mu\nu}(x)G^{\mu\nu}(x)\right )+\bar
q(x)(i\hat{D}-M)q(x)$.\\[9pt]
 $M$ mixes  left and right spaces
 $q_L(x)$ and $q_R(x)$.
 But in  chiral limit  $M\to 0$  these spaces
 separate realizing    $U_L(3)\times U_R(3)$
 symmetry  accurate within violation through gluonic anomaly.\\[6pt]
 As
 Experiment suggests,  Confinement forms colourless
observable hadronic fields and spontaneous breaking of chiral
symmetry with massless pseudoscalar fields.\\[6pt]
 There are two
possible scenarios for  QCD at low energy.\\[6pt]
  1.
$U_L(3)\times U_R(3)$ non-linear $\sigma$-model.\\[3pt]
  2. $U_L(3)\times U_R(3)$ linear
$\sigma$-model.\\
 The experimental nonet of the light scalar
mesons suggests   $U_L(3)\times U_R(3)$ linear  $\sigma$-model.

\section{History of light scalar mesons}

 Hunting   the light  $\sigma$ and  $\kappa$
mesons had begun in the sixties already. But long-standing
unsuccessful attempts to prove their existence in a  conclusive
way entailed general disappointment and a preliminary information
on these states disappeared from Particle Data Group (PDG)
Reviews. One of principal reasons against the  $\sigma$ and
$\kappa$ mesons was the fact that both  $\pi\pi$ and  $\pi K$
scattering phase shifts  do not pass over  $90^0$ at putative
resonance masses \cite{narrow}.

\section{\boldmath{$ SU_L(2)\times SU_R(2)$} linear $\sigma$ model \cite{GML,AS94,AS07}}

  Situation  changes when we showed (1994) that in the
linear  $\sigma$ model
\begin{eqnarray}
&&  L=\frac{1}{2}\left
[(\partial_\mu\sigma)^2+(\partial_\mu\overrightarrow{\pi})^2\right
]- \frac{m_\sigma^2}{2}\sigma^2- \frac{m_\pi^2}{2}
\overrightarrow{\pi}^2 \nonumber\\[9pt]
 && -
\frac{m_\sigma^2-m_\pi^2}{8f^2_\pi}\left [\left
(\sigma^2+\overrightarrow{\pi}^2\right )^2 +4f_\pi\sigma \left
(\sigma^2+\overrightarrow{\pi}^2\right ) \right ]^2\nonumber
\end{eqnarray}
 there is a  negative background
phase which  hides the  $\sigma$ meson \cite{AS94}. It has been
made clear that  shielding wide lightest scalar mesons in chiral
dynamics is very  natural.

 This idea was picked up and triggered
new wave of theoretical and experimental searches for the
 $\sigma$ and  $\kappa$ mesons.

\subsection{Our approximation \cite{AS94,AS07}}
\begin{figure}[h]
  \begin{center}
    \includegraphics[width=\textwidth,height=7cm]{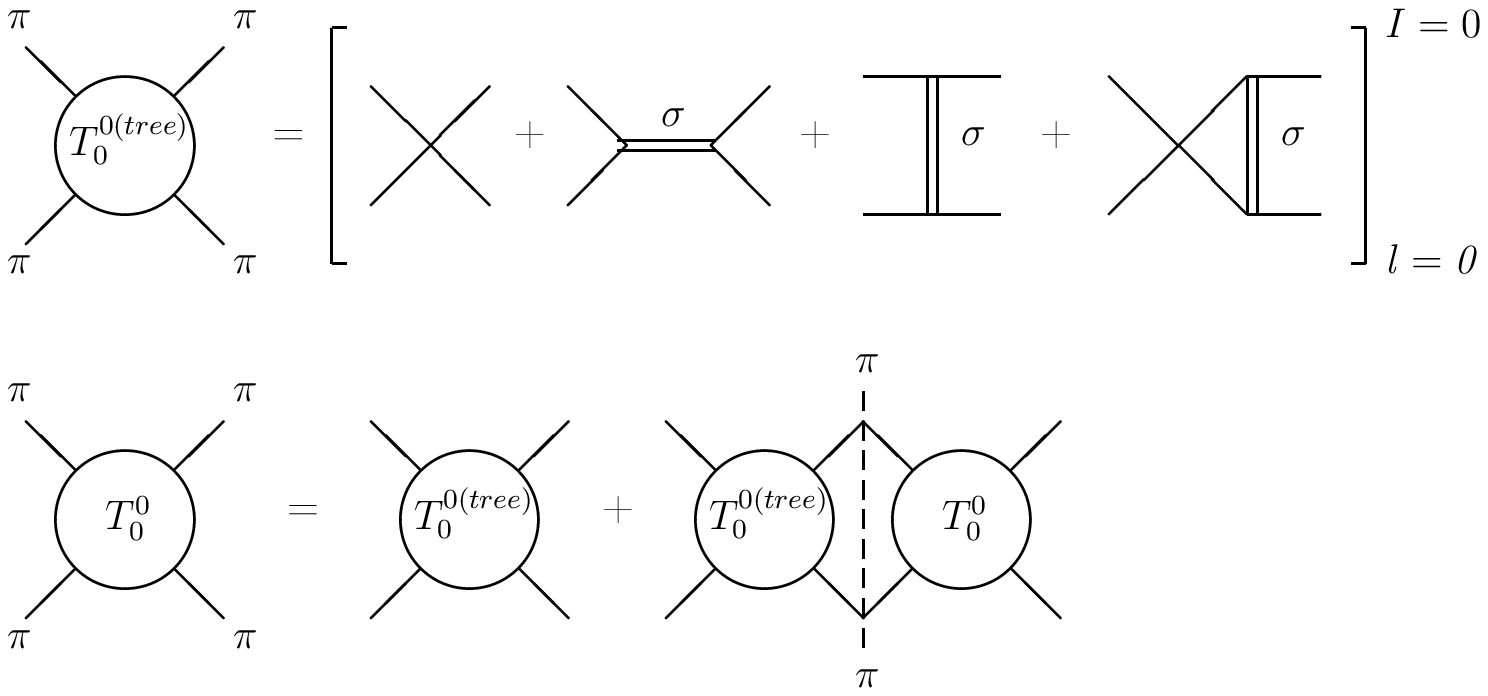}
    \caption{Our approximation.}
    \label{fig:Our approximation}
  \end{center}
\end{figure}

\subsection{Chiral shielding in  \boldmath{$\pi\pi\to\pi\pi$} in
our approximation  \cite{AS94,AS07}}

$m_\sigma = 0.93$ GeV,\ \ $M_{res}=0.43$ GeV,\ \
$\Gamma^{renorm}_{res}(M^2_{res})=0.53$ GeV

\begin{figure}[h]
  \begin{center}
    \includegraphics[width=\textwidth,height=8cm]{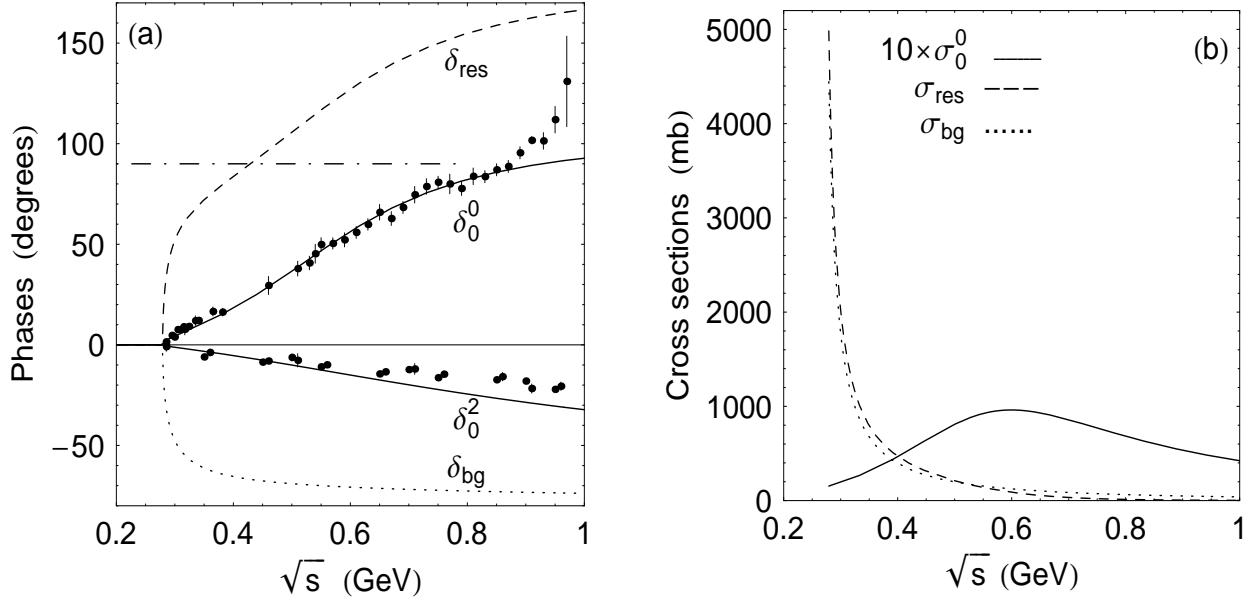}
    \caption{The  $\sigma$ model.  Our approximation. $\delta^0_0=\delta_{res}+\delta_{bg}$.}
    \label{fig:ChirShiel}
      \end{center}
     \end{figure}

\newpage

\subsection{The \boldmath{$\sigma$}  pole in $\pi\pi\to\pi\pi$ in
our approximation  \cite{AS07}}

\begin{eqnarray}
 &&
T^0_0\to\frac{g^2_\pi}{s-s_\sigma}\,,\ \ \ g^2_\pi=(0.12+i0.21)\,
\mbox{GeV}^2\,,\nonumber\\[9pt]
 &&  s_\sigma=(0.21-i0.26)\,\mbox{GeV}^2\,,\ \ \
\sqrt{s_\sigma}=M_\sigma-i\frac{\Gamma_\sigma}{2}=(0.52-i0.25)\,\mbox{GeV}.\nonumber
\end{eqnarray}
Considering the residue of the  $\sigma$ pole in $T^0_0$ as the
square of its coupling constant to the $\pi\pi$ channel is not a
clear guide to understand the  $\sigma$ meson nature for its great
obscure imaginary part.

\subsection{The \boldmath{$\sigma$} propagator in our approximation
\cite{AS94,AS07}}

Another thing is the $\sigma$ propagator

\begin{equation}
 \frac{1}{D_\sigma (s)}= \frac{1}{M^2_{res} - s
+ \mbox{Re}\Pi_{res}(M^2_{res})- \Pi_{res}(s)}\,.\nonumber
\end{equation}
 The $\sigma$ meson self-energy $\Pi_{res}(s)$ is caused
by the intermediate $\pi\pi$ states, that is, by  the four-quark
intermediate states. This contribution shifts the Breit-Wigner
(BW) mass greatly $m_\sigma - M_{res}= 0.50\,\mbox{GeV}$. So, half
the BW mass is determined by
 the four-quark contribution  at least. The imaginary part
dominates the propagator modulus in the region 300 MeV $<
\sqrt{s}<$ 600 MeV. So, the $\sigma$ field is described by  its
four-quark component  at least in this energy (virtuality)
 region.

\section{Troubles and expectancies}

In theory the  principal problem is  impossibility to use the
linear  $\sigma$ model in the  tree level approximation inserting
widths into  $\sigma$ meson propagators because such an approach
 breaks the both  unitarity and the  Adler
self-consistency condition. The  comparison with the experiment
 requires the  non-perturbative calculation of
the process amplitudes.  Nevertheless, now there are the
possibilities to estimate  odds of the $U_L(3)\times U_R(3)$
linear $\sigma$ model to  underlie physics of light scalar mesons
 in phenomenology, taking into account  the idea of
chiral shielding of  $\sigma (600)$,  our treatment of
$\sigma(600)$-$f_0(980)$ mixing based on quantum field theory
ideas,  Adler's condition, and results, obtained on the base of
the chiral expansion and the Roy equations \cite{CCL}.

\section{Phenomenological treatment \cite{AK11}, \boldmath{$\delta_0^0=\delta_B^{\pi\pi} +
\lowercase{\delta_{res}$}}}

\begin{figure}[h]
  \begin{center}
    \includegraphics[width=\textwidth,height=7
    cm]{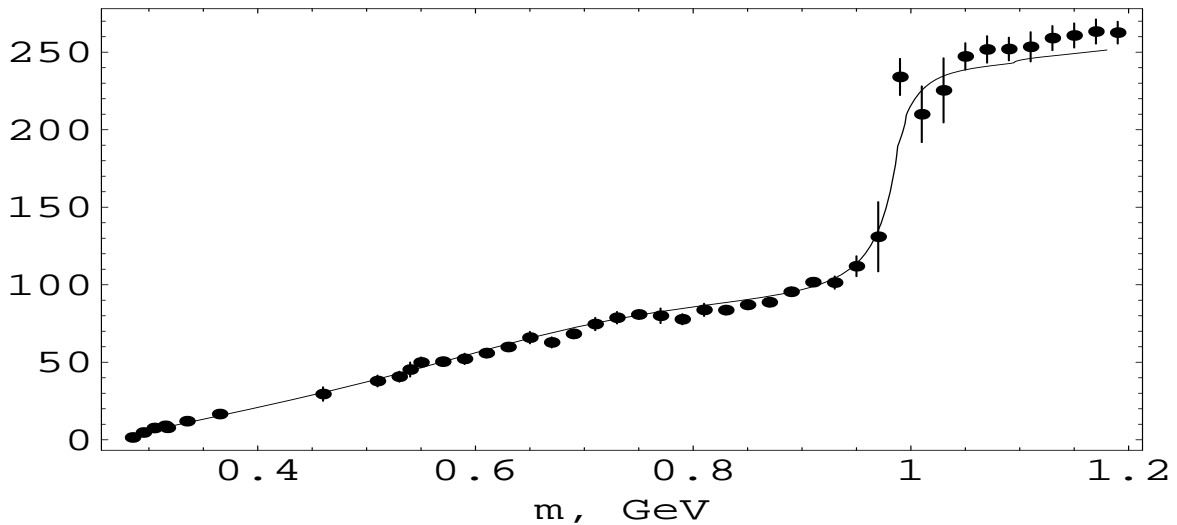}
    \caption{$\delta_0^0=\delta_B^{\pi\pi} + \delta_{res}$.}
    \label{fig: pipiphase}
      \end{center}
    \end{figure}

\begin{figure}[h]
  \begin{center}
    \includegraphics[width=\textwidth,height=6cm]{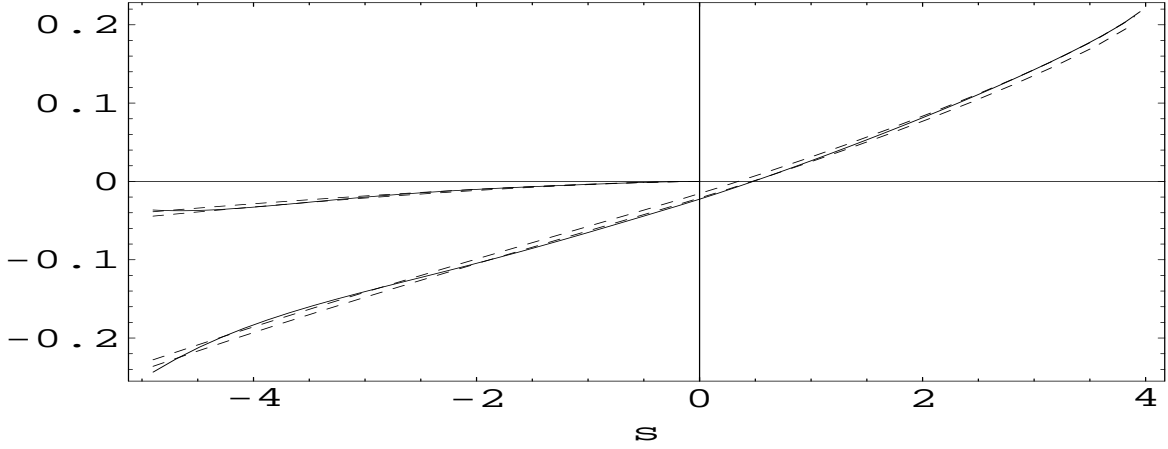}
    \caption{$T^0_0$ \cite{AK11}, comparison with the calculations based on the chiral expansion and the Roy equations  \cite{CCL},
    $s$ in units of $m_\pi^2$\,;\ \ \ the
real part under the threshold: $-5< s< 4$\,; \ \ \ the imaginary
part on the left cut: $-5<s<0$.}
    \label{fig: CCL}
      \end{center}
    \end{figure}
    \vspace*{-2cm}

    \begin{eqnarray}
&& g_{\sigma\pi^+\pi^-}^2/4\pi=0.57\, \mbox{ GeV}^2,\quad
g_{\sigma K^+K^-}^2/4\pi=0.048\, \mbox{ GeV}^2\nonumber\\[6pt]
 &&
g_{f_0\pi^+\pi^-}^2/4\pi=0.36\, \mbox{ GeV}^2,\quad
g_{f_0K^+K^-}^2/4\pi=2\, \mbox{GeV}^2\nonumber\\[6pt]
 &&
 m_\sigma=507\,\mbox{ MeV}\,,\
 \Gamma_\sigma (m_\sigma)=353\,\mbox{ MeV}\,,\
 m_{f_0}=987\,\mbox{MeV}\,,\nonumber\\[6pt]
&& \Gamma_{f_0} (m_{f_0})=130\,\mbox{ MeV}\,,\ \ \ a^0_0=0.226\,
m^{-1}_{\pi^+}\nonumber
\end{eqnarray}

\subsection{The \boldmath{$\sigma(600)$} poles \cite{AK11}}
\begin{table}[h]
\begin{center}
\begin{tabular}{|c|c|c|c|c|c|}\hline
$\Pi^{\pi\pi}(s)$  & $\Pi^{K\bar K}(s)$  & $\Pi^{\eta\eta}(s)$  &
$\Pi^{\eta\eta'}(s)$  & $\Pi^{\eta'\eta'}(s)$   & Fit
\\ \hline

II & I & I & I & I &  $613.8 - 221.4\,i$
\\ \hline

II & II & I & I & I  & $609.8 - 291.6\,i$
\\ \hline

II & II & II & I & I  & $559.4 - 346.6\,i$
\\ \hline

II & II & II & II & I  & $569.7 - 410.7\,i$
\\ \hline

II & II & II & II & II  & $581.6 - 411.0\,i$
\\ \hline
\end{tabular}
\caption{The $\sigma(600)$ poles (MeV) on different sheets of the
complex $s$
 plane depending on lists of polarization operators
$\Pi^{ab}(s)$.}
 \label{tab:sigmapoles}
\end{center}
\end{table}

I. Caprini, G. Colangelo, and H. Leutwyler  \cite{CCL}:\\
$\sqrt{s_\sigma}= M_\sigma-i\Gamma_\sigma/2 = (
441^{+16}_{-8}-i272^{+9}_{-12.5} )\times$ MeV\,.\\[9pt]
    But
the Roy equations are approximate, they take into account only the
$\pi\pi$ channel, but the true $\pi\pi$ scattering amplitude has
the multi (infinity)-list Riemannian  surface, that can effect the
analytical continuation considerably, especially in the  wide
resonance case.

 So, the current activity, aiming extremely precise determination of
the $\sigma(600)$ pole position, has taken the forms of the
Swift's grotesque. Really, the residue of the $\sigma$ pole can
not be connected to coupling constant in the Hermitian (or
quasi-Hermitian), see Subsection IV.C and Ref. \cite{AS07}, for it
has a large imaginary part and this pole can not be interpreted as
a physical state for its huge width.

\subsection{The \boldmath{$f_0(980)$} poles \cite{AK11}}
\begin{table}[h]
 \begin{center}
\begin{tabular}{|c|c|c|c|c|c|}\hline

$\Pi^{\pi\pi}(s)$ & $\Pi^{K\bar K}(s)$  & $\Pi^{\eta\eta}(s)$  &
$\Pi^{\eta\eta'}(s)$  & $\Pi^{\eta'\eta'}(s)$   & Fit  \\ \hline

II & I & I & I & I &  $990.5 - 19.4\,i$
\\ \hline

II & II & I & I & I  & $1183.2 - 518.6\,i$
\\ \hline

II & II & II & I & I  & $1366.0 - 756.5\,i$
\\ \hline

II & II & II & II & I & $1390.7 - 813.0\,i$
\\ \hline

II & II & II & II & II  & $1495.6 - 1057.7\,i$
\\ \hline
\end{tabular}
\caption{The $f_0(980)$ poles (MeV) on different sheets of the
complex $s$
 plane depending on lists of polarization operators
$\Pi^{ab}(s)$.}
 \label{tab:f0poles}
\end{center}
\end{table}

The futility of the approach that is based on the poles treatment
may be additionally illustrated by Fit. As seen on line 1 of Table
II, the real part of the $f_0(980)$ pole $Re M_{f_0}$ on the II
sheet of the $T^0_0$ exceeds the $K^+K^-$ threshold ($987.4$ MeV),
it means that $Im M_{f_0}$ equals to
$-(\Gamma(f_0\to\pi\pi)-\Gamma(f_0\to K^+K^-))/2$, which is
physically meaningless.  In this case we should take $\Pi^{K^+
K^-}$ on the second sheet, this gives the pole at $M_{f_0}=( 989.6
-168.7i$)\,MeV, with $Re M_{f_0}$ between the $K^+K^-$ and
$K^0\bar{K^0}$ thresholds again. But, the analytical properties
are specified  on the  $s$ plane, and we must consider not
$M_{f_0}$, but $M_{f_0}^2 = (0.951 - 0.334\,i)$ GeV$^2$. So, we
have the pole with a real part below the $K^+K^-$  (0.975 GeV$^2$)
and $K^0\bar{K^0}$ thresholds, and an imaginary part dictated by
analytical continuation of the kaon polarization operators.

\section{Four-quark model \cite{Ja,1984,1998}}

The  nontrivial nature of the well-established light scalar
resonances $f_0(980)$ and  $a_0(980)$ is no longer denied
practically anybody. As for the nonet as a whole, even a cursory
look at PDG Review gives an idea of the  four-quark structure of
the light scalar meson nonet, $\sigma (600)$, $\kappa (800)$,
$f_0(980)$, and $a_0(980)$, inverted in comparison with the
classical $P$ wave $q\bar q$ tensor meson nonet, $f_2(1270)$,
$a_2(1320)$, $K_2^\ast(1420)$, $\phi_2^\prime (1525)$. Really,
while the scalar nonet  cannot be treated as the $P$ wave $q\bar
q$ nonet in the naive
 quark model, it can be
easy understood as the $q^2\bar q^2$ nonet, where $\sigma$ has no
strange quarks, $\kappa $ has the  s quark, $f_0$ and $a_0$ have
the $s\bar s$ pair. Similar states were found by Jaffe in 1977 in
the MIT bag \cite{Ja}.

\section{Radiative decays of \boldmath{$\phi$}-meson \cite{AI}}

Ten years later (1987, 1989) we showed that $\phi\to\gamma
a_0\to\gamma\pi\eta$ and $\phi\to\gamma f_0\to \gamma\pi\pi$ can
shed light on the problem of $a_0(980)$ and $f_0(980)$ mesons
\cite{AI}.   Now these decays are studied not only theoretically
but also experimentally. The first measurements  (1998, 2000) were
reported by  SND and  CMD-2.   After (2002) they were studied  by
KLOE  in agreement with the Novosibirsk data
 but with a considerably smaller error.

  Note that $a_0(980)$  is produced in the
radiative $\phi$ meson decay
 as intensively as  $\eta '(958)$  containing  $\approx 66\% $ of $s\bar s$,
  responsible for  $\phi\approx s\bar s\to\gamma s\bar s\to\gamma \eta '(958)$.
 It is a clear qualitative argument
for the presence of the  $s\bar s$ pair in the isovector
$a_0(980)$ state, i.e., for its  four-quark nature.
 \vspace*{-1cm}
\subsection{The \boldmath{$K^+K^-$}-loop model \cite{AI,AG9701,AK03,A03,AK06,AK11}}
\begin{figure}[h]
\begin{center}
\vspace*{0.5cm}
\begin{tabular}{ccc}
\hspace*{-24pt}
\includegraphics[width=3.3cm]{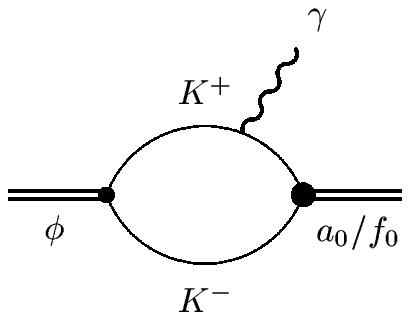}&
\raisebox{-6mm}{$\includegraphics[width=3.3cm]{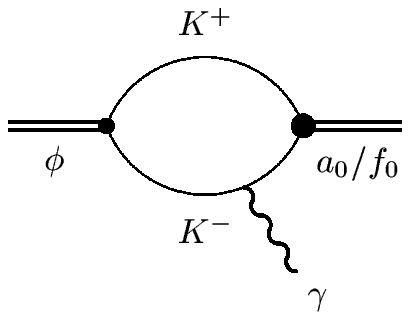}$}&
\includegraphics[width=3.3cm]{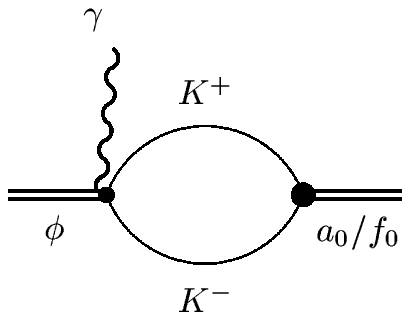}\\ (a)&(b)&(c)
\end{tabular}
\caption{The kaon loop model \cite{AI}.}
    \label{fig: kloop}
\end{center}
\end{figure}

\vspace*{-0.5cm}
 When basing the experimental investigations,
we suggested \cite{AI} the kaone-loop model $\phi\to
K^+K^-\to\gamma (a_0/f_0)$, Fig. \ref{fig: kloop}.
 \vspace*{-0.5cm}
\newpage

 This model is
used in the data treatment and is ratified by experiment.
\begin{figure}[h]
  \begin{center}
    \includegraphics[width=\textwidth,height=8cm]{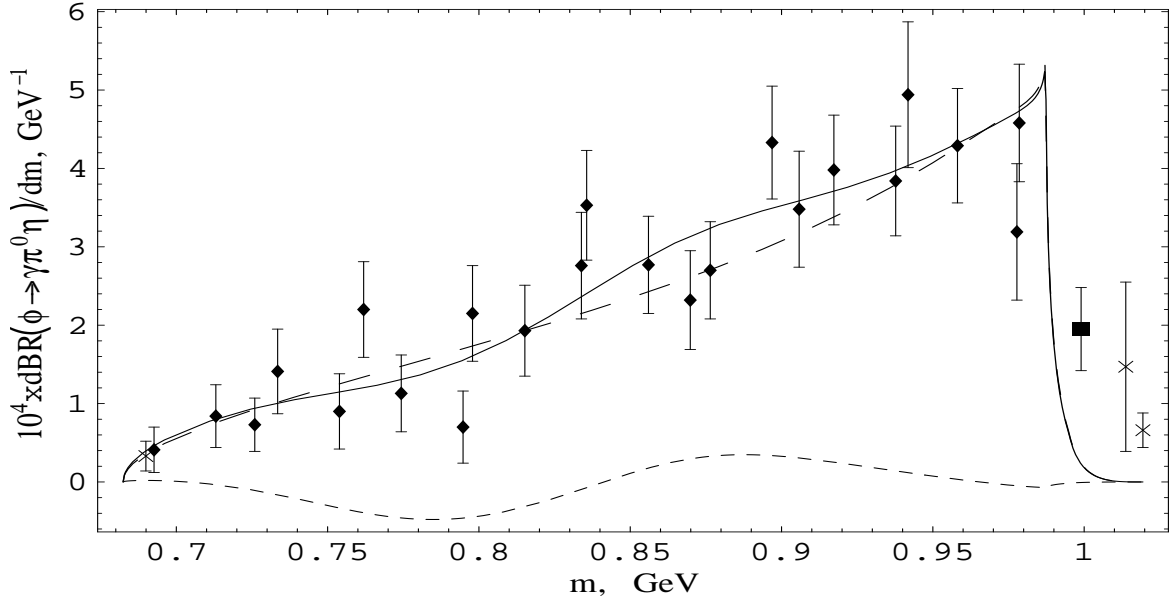}
    \caption{The KLOE data on $\phi\to\gamma\pi^0\eta$\,, the theory \cite{AK03}.}
    \label{fig: phitogammaa0}
      \end{center}
    \end{figure}

\begin{figure}[h]
  \begin{center}
    \includegraphics[width=\textwidth,height=8cm]{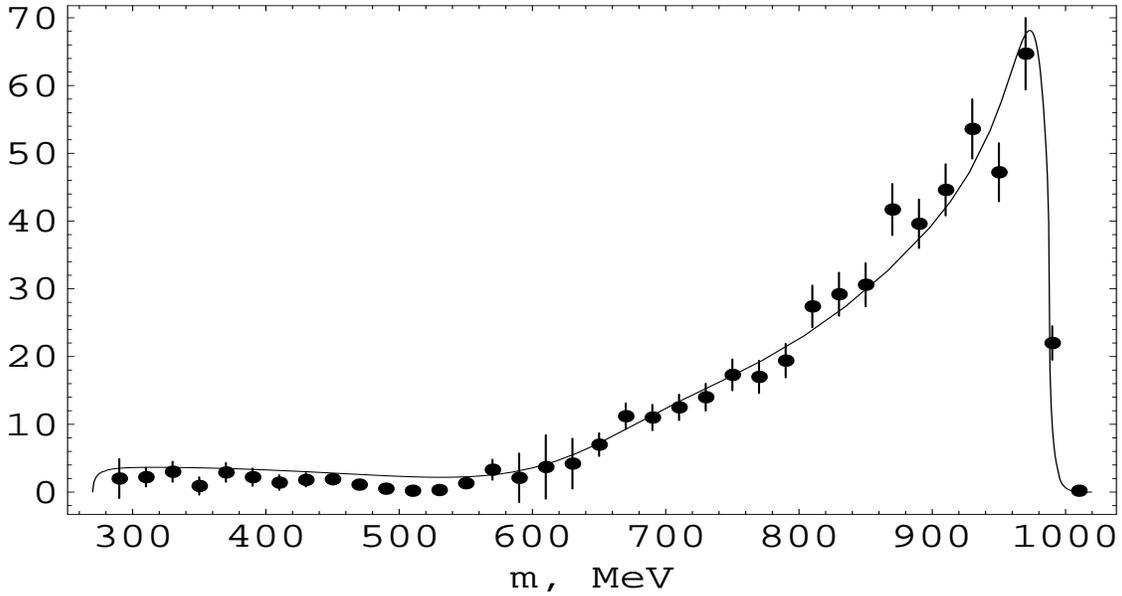}
    \caption{The KLOE data on $\phi\to\gamma\pi^0\pi^0$\,, the theory \cite{AK11}.}
    \label{fig: phitogammaf0}
      \end{center}
    \end{figure}

 Gauge invariance gives the conclusive
arguments in favor of the $K^+K^-$ - loop transition as the
principal mechanism of $a_0(980)$ and $f_0(980)$ production in the
$\phi$ radiative decays \cite{A03}.

The point is to describe the experimental spectra one should to
stop the rapid growth of the $\omega(m)^3$ function, arising of
gauge invariance and the phase space factor, at
$\omega(990\,\mbox{MeV})=29$ MeV, where $\omega(m)$ is the photon
energy.

 The
 $K^+K^-$-loop model  $\phi\to K^+K^-\to\gamma
R$ solves this problem in  the elegant way with the help of the
nontrivial  threshold phenomenon.

 So, the
mechanism of production of $a_0(980)$ and $f_0(980)$ mesons in the
$\phi$ radiative decays is established  at a physical level of
proof:  {\bf WE ARE DEALING WITH THE FOUR-QUARK TRANSITION}.

 A radiative four-quark transition between two  $q\bar
q$ states requires creation and annihilation of an additional
 $q\bar q$ pair, i.e., such a transition is forbidden
according to the OZI rule, while a radiative four-quark transition
between  $q\bar q$ and  $q^2\bar q^2$ states requires only
creation of an additional  $q\bar q$ pair, i.e., such a transition
is allowed according to the  OZI rule.  The large
 $N_C$ expansion supports this conclusion \cite{A03}.

 \section{ \boldmath{\lowercase{$a_0(980)/f_0(980)\to\gamma\gamma$}} and $q^2\bar q^2$-model \cite{ADS}}

  Twenty nine years ago (1982) we predicted the suppression of
$a_0(980)\to\gamma\gamma$ and $f_0(980)\to\gamma\gamma$ in the
$q^2\bar q^2$ MIT model,
$\Gamma(a_0(980)\to\gamma\gamma)\sim\Gamma(f_0(980)\to\gamma\gamma)\sim
0.27\,\mbox{keV}$ \cite{ADS}.

 Experiment supported this prediction
\begin{eqnarray}
 &&\Gamma (a_0\to\gamma\gamma)=(0.19\pm 0.07
^{+0.1}_{-0.07})/B(a_0\to\pi\eta) \, \mbox{keV, Crystal
Ball}\nonumber\\
 && \Gamma (a_0\to\gamma\gamma)=(0.28\pm 0.04\pm
0.1)/B(a_0\to\pi\eta)\, \mbox{keV, JADE.}\nonumber\\
&&\Gamma
(f_0\to\gamma\gamma)=(0.31\pm 0.14\pm 0.09)\, \mbox{keV, Crystal
Ball,}\nonumber\\
&&\Gamma (f_0\to\gamma\gamma)=(0.24\pm 0.06\pm
0.15)\, \mbox{keV, MARK II}.\nonumber
\end{eqnarray}
 When in the $q\bar q$ model it was anticipated
\begin{eqnarray}
 &&\Gamma(a_0\to\gamma\gamma)=(1.5 - 5.9)\Gamma
(a_2\to\gamma\gamma) = (1.5 - 5.9)(1.04\pm 0.09)\,\mbox{keV.}
\nonumber\\
 &&\Gamma(f_0\to\gamma\gamma)=(1.7 - 5.5)\Gamma
(f_2\to\gamma\gamma)  = (1.7 - 5.5)(2.8\pm
0.4)\,\mbox{keV.}\nonumber
\end{eqnarray}

\section{ Nature of light scalar mesons and their production mechanisms in \boldmath{$\gamma\gamma$}
collisions \cite{AS88,Bel07,Bel08,Bel09,AS08,AS10,AS11}}

Recently  the experimental investigations have made great
qualitative advance. The Belle Collaboration  published data on
 $\gamma\gamma\to\pi^+\pi^-$ (2007),
$\gamma\gamma\to\pi^0\pi^0$ (2008), and $\gamma\gamma\to\pi^0\eta$
(2009), whose statistics are  huge. They not only proved the
theoretical expectations based on the four-quark nature of the
light scalar mesons,  but also have allowed to elucidate the
principal mechanisms of these processes.

 Specifically, the direct
coupling constants of the $\sigma(600)$, $f_0(980)$, and
$a_0(980)$resonances with the $\gamma\gamma$ system are small with
the result  that their decays in the two photon are the four-quark
transitions caused by the rescatterings $\sigma\to\pi^+\pi^-\to
\gamma\gamma$, $f_0(980)\to K^+K^-\to\gamma\gamma$, and
$a_0(980)\to K^+K^- + \pi^0\eta\to\gamma\gamma$ in contrast to the
two-photon decays of the classic $P$ wave tensor $q\bar q$ mesons
$a_2(1320)$, $f_2(1270)$ and $f'_2(1525)$, which are caused by the
direct two-quark transitions  $q\bar q\to\gamma\gamma$ in the
main. As a result the practically model-independent prediction of
the $q\bar q$ model
$g^2_{f_2\gamma\gamma}:g^2_{a_2\gamma\gamma}=25:9$ agrees with
experiment rather well.

The two-photon light scalar widths averaged over resonance mass
distributions $\langle\Gamma_{f_0\to\gamma
\gamma}\rangle_{\pi\pi}$ $\approx$\,0.19 keV,
$\langle\Gamma_{a_0\to\gamma \gamma}\rangle_{\pi\eta}\approx
$\,0.3 keV and $\langle\Gamma_{\sigma
\to\gamma\gamma}\rangle_{\pi\pi}\approx$\,0.45 keV.  As to the
ideal  $q\bar q$ model prediction
$g^2_{f_0\gamma\gamma}:g^2_{a_0\gamma\gamma}=25:9$, it is excluded
by experiment.

\begin{figure}[h]
  \begin{center}
  \includegraphics[width=\textwidth,height=8cm]{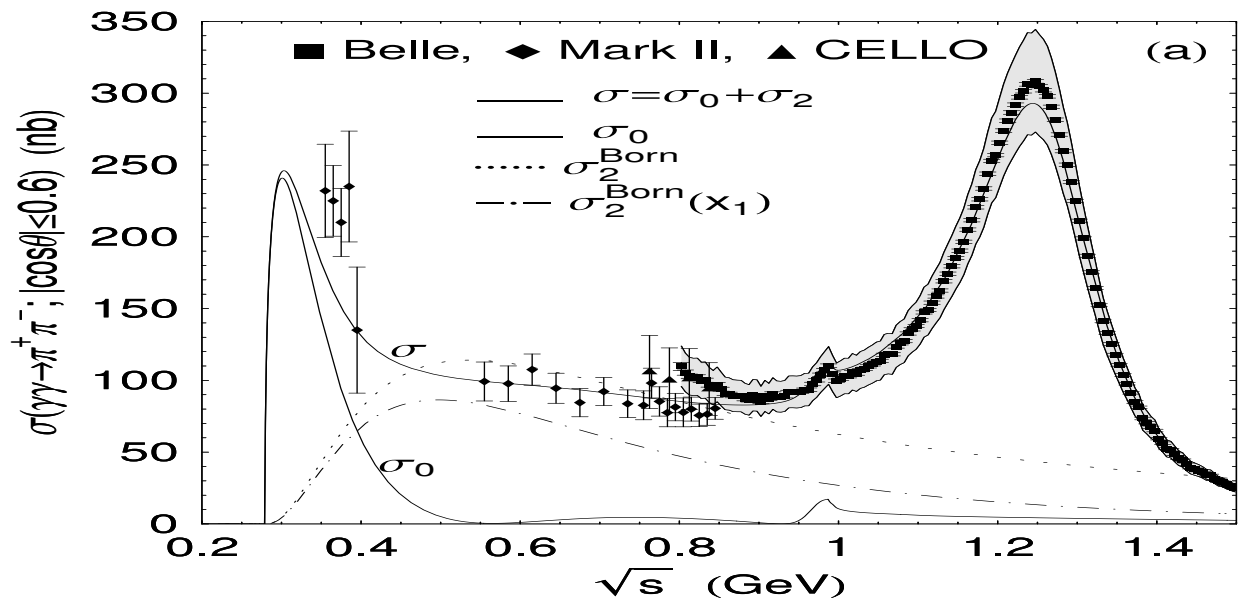}
    \caption{The Belle data on $\gamma\gamma\to\pi^+\pi^-$ \cite{Bel07}, the theory  \cite{AS08,AS11}.}
    \label{fig:bellepi+pi-}
  \end{center}
\end{figure}

\begin{figure}[h]
  \begin{center}
  \includegraphics[width=\textwidth,height=8cm]{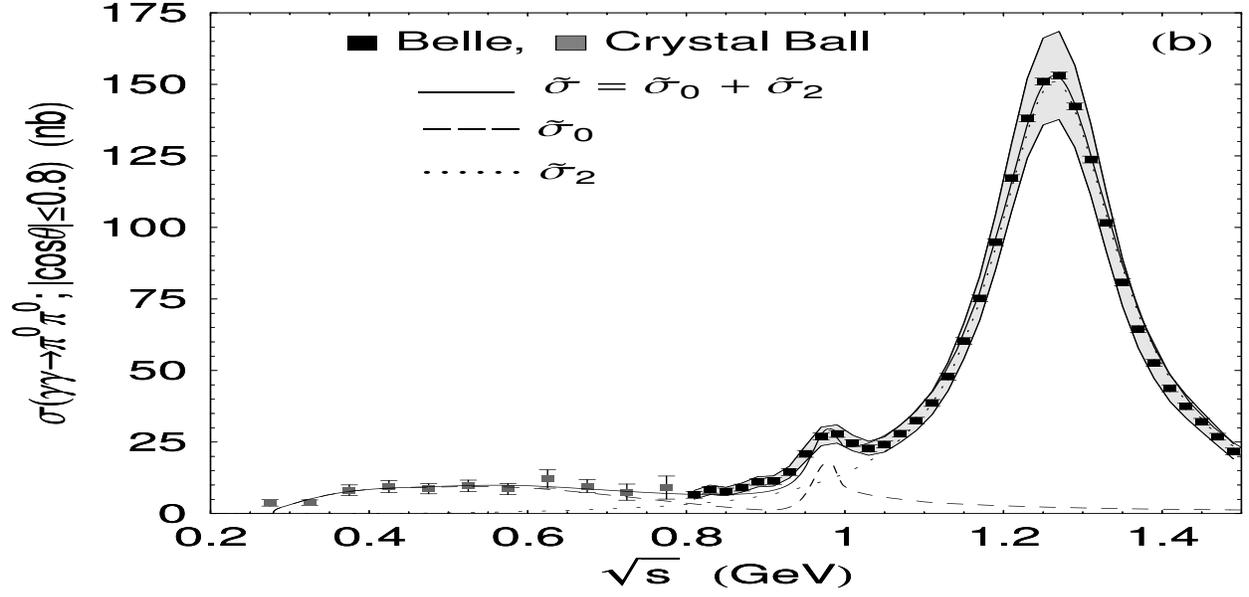}
    \caption{The Belle data on $\gamma\gamma\to\pi^0\pi^0$ \cite{Bel08}, the theory \cite{AS08,AS11}.}
    \label{fig:bellepi0pi0}
  \end{center}
\end{figure}

\begin{figure}[h]
  \begin{center}
  \includegraphics[width=\textwidth,height=8cm]{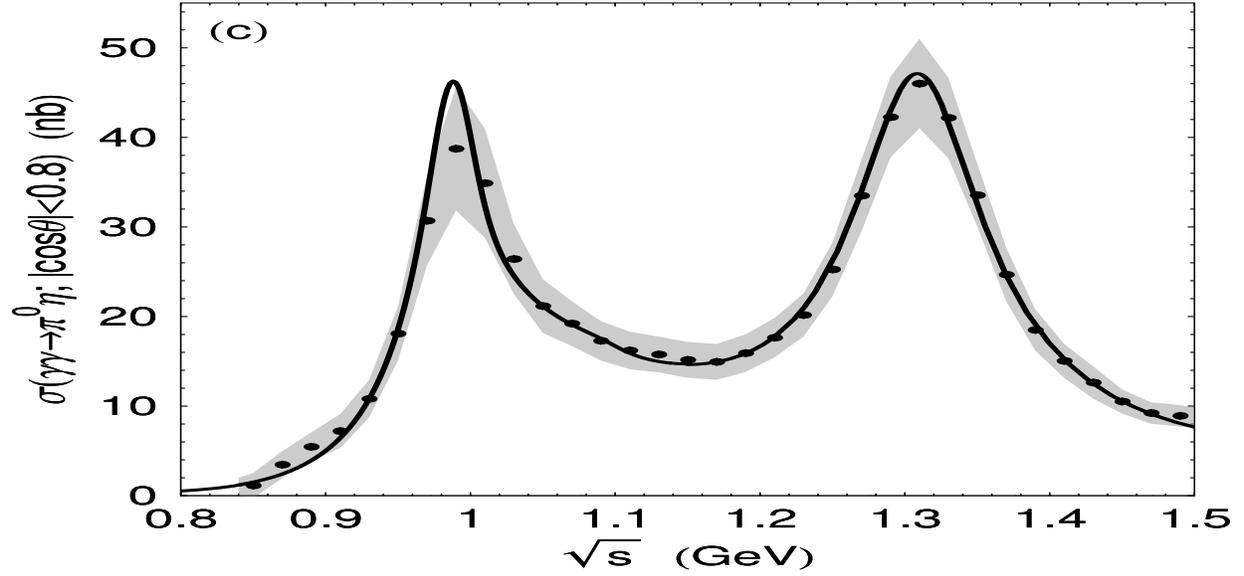}
    \caption{The Belle data on $\gamma\gamma\to\pi^0\eta$ \cite{Bel09}, the theory \cite{AS10,AS11}.}
    \label{fig:bellepi0eta}
  \end{center}
\end{figure}

\section{Lessons}

The mass  spectrum of the light scalars,    $\sigma (600)$,
$\kappa (800)$, $f_0(980)$, $a_0(980)$,  gives an idea of their
$q^2\bar q^2$ structure.

 Both intensity and mechanism of the
$a_0(980)/f_0(980)$ production in the radiative decays of
$\phi(1020)$, the  $q^2\bar q^2$ transitions $\phi\to
K^+K^-\to\gamma [a_0(980)/f_0(980)]$, indicate their $q^2\bar q^2$
nature.

 Both intensity and mechanism of the scalar
meson decays into $\gamma \gamma$, basically the four-quark
transitions, $\sigma(600)\to\pi^+\pi^-\to\gamma\gamma$, $
f_0(980)\to K^+K^-\gamma\gamma$, and $a_0(980)\to K^+K^- +
\pi^0\eta\to\gamma\gamma$, indicate their $q^2\bar q^2$
nature,too.

In addition, the  absence of the $J/\psi$ $\to$ $\gamma f_0(980)$,
$ a_0(980)\rho$, $f_0(980)\omega$ decays in contrast to the
intensive the $J/\psi$ $\to$ $\gamma f_2(1270)$,$ \gamma
f'_2(1525)$, $a_2(1320)\rho$, $f_2(1270)\omega$ decays  intrigues
against the $P$ wave  $q\bar q$ structure of $a_0(980)$ and
$f_0(980)$ also.

\section{Outlook}

\begin{enumerate}
 \item $\gamma\gamma\to K^+K^-\,,\ K^0\overline{K^0}$ near the thresholds,
 it is expected a  drastic suppression of the  Born contribution in the $K^+K^-$ channel.
 $\gamma\gamma^*(Q^2)\to\pi^0\pi^0\,,\ \pi^0\eta$,
it is expected a drastic decrease of the $\sigma(600)$, $f_0(980)$
and $a_0(980)$ contributions  with increasing $Q^2$ as opposed to
a decrease of the $f_2(1270)$ and $a_2(1320)$ ones.
\item Search for $J/\psi\to f_0(980)\omega$ and $J/\psi\to
  a_0(980)\rho$\,.
 \item  Search for the $a_0(980)-f_0(980)$ mixing \cite{ADS79} in\\
 i) $J/\psi\to
  f_0(980)\phi\to
  a_0(980)\phi\to\pi^0\eta\phi$ \cite{Zou1,Zou2} and\\
 ii)   $\pi^- p\to  f_0(980)n\to a_0(980)n\to\pi^0\eta
n$ \cite{AS04}\,,\\ here
 it is expected a  strong jump in the spin asymmetry that could give
  an exclusive information on  ($g_{a_0K^+K^-}\cdot g_{f_0K^+K^-})
  /4\pi$.
\item The new precise experiment on
$\pi\pi\to K\bar K$ would give the crucial information about the
inelasticity $\eta^0_0$ and about the phase $\delta_B^{K\bar
K}(m)$ near the $K\bar K$ threshold.  The precise measurement of
the inelasticity $\eta^0_0$ near 1 GeV in $\pi\pi\to\pi\pi $ would
also be very important.
\end{enumerate}

 \acknowledgements{I am grateful to G.N. Shestakov,
S.A. Devaynin, V.N. Ivanchenko, V.V. Gubin, and A.V. Kiselev who
helped me to learn these lessons.

 This work was supported in part by RFBR,
Grant No 10-02-00016.}


\begin{thebibliography}{99}
\bibitem{narrow} Meanwhile, the narrow light scalar,
$a_0(980)$ and   $f_0(980)$, were discovered.
\end{thebibliography}
\end{document}